# A simple, reliable, and no-destructive method for the measurement of vacuum pressure

Jinpeng Yuan・Zhonghua Ji・Yanting Zhao・Xuefang Chang・Liantuan Xiao・Suotang Jia

*State Key Laboratory of Quantum Optics and Quantum Optics Devices, Laser Spectroscopy Laboratory, Shanxi University, Taiyuan 030006, China*

**Abstract** We present a simple, reliable, and no-destructive method for the measurement of vacuum pressure in a magneto-optical trap. The vacuum pressure is verified to be proportional to collision rate constant between cold atoms and background gas with a coefficient k, which can be calculated by simple ideal gas law. The rate constant for loss due to collisions with all background gases can be derived from the total collision loss rate by a series of loading curve of cold atoms under different trapping laser intensities. The presented method is also applicable for other cold atom systems and meets the miniaturization requirement of commercial applications.

## 1 Introduction

With the development of laser cooling and trapping of neutral atoms, atom physics has promised a variety of research directions, such as time and frequency metrology, atom interferometry and inertial sensors, atom lasers, simulation of condensed matter systems, production and study of strongly correlated systems, and production of ultra-cold molecules [1]. Since its invention, the magneto-optical trap (MOT) [2] has become the most important primary device for preparation of a sample of cold atoms. It is now widely appreciated for fundamental discoveries and practical applications in such diverse fields as quantum degenerate gases, cold collisions, quantum information processing, ultraprecise frequency standards, quantum optics, and trace atom detection [3].

In forming a MOT, one of the most important and widely considered fundamental requirements is the generation and maintenance of vacuum pressure. For these two aspects, they are all inseparable from the measurement of vacuum pressure [4].

The standard instrument for vacuum pressure measurement is the ionization gauge [5]. The vacuum pressure can be gotten indirectly by the measurement of the positive ion current generated from the colliding between the gas molecules and charged particles. The instrument can measure pressures to $10^{-9}$ Pa or lower. However, ionization gauge requires typically 100 W of electrical power and take up volume of 100 cm$^3$ or more. The requirement may be negligible in large laboratory-based vacuum systems. However, as systems are miniaturized and streamlined to improve simplicity and efficiency for business applications or some special applications, the ionization gauge is likely to become unacceptable. Another measurement instrument is the residual gas analyzer [6]. Based on the relation between vacuum pressure and the composition of the residual gas generated by the manufacture and preparation process, the accurate analysis of the residual gas can be a reliable method to measure the vacuum pressure. But it also suffers from similar constraints. Besides of the two equipments, the ion pump is also used to measure the vacuum pressure while synchronously maintain the vacuum pressure [7]. The working current of background gas under high voltage reflects the vacuum pressure value. However, the ion pumps are not widely used to measure vacuum pressure because of the leakage currents limit the minimum pressure reading. The lower accuracy which arises from the complex relation between of ion pump pressure and the working current is also a drawback to measure vacuum pressure. In addition, the value of vacuum pressure measured by ion pump is far away from the cavity where the cold atoms exist.

In order to miniaturize the vacuum pressure measurement of cold atoms device, several groups has investigated the relation between vacuum pressure and the cold atoms collision loss rate with background gas. M. Prentiss *et al.* [8] found the collision loss rate varies inversely with the background pressure and insensitive to the depth of the trap in a magnetic-molasses optical trap. M. H. Anderson *et al.* [9] investigated the relationship between the collision loss rate and the background pressure in MOT and forced dark-spot trap. P. A. Willems *et al.* [10] use the collision loss rate as a method to estimate the background pressure in a quadrupole magnetostatic trap. They use the known He-Cs van der Waals

J. P. Yuan・Z. H. Ji・Y.T. Zhao (✉)・X. F. Chang・L. T. Xiao・S. T. Jia
State Key Laboratory of Quantum Optics and Quantum Optics Devices, Laser Spectroscopy Laboratory, Shanxi University, Taiyuan 030006, China
E-mail: zhaoyt@sxu.edu.cn, Fax: +86-351-7011645

collision cross section to infer typical background gas pressures to be below $4\times10^{-12}$ Torr. They all found that the background pressure decreases as the collision loss rate increases. They did not calculate the relationship between them and only gave a quantitative expression. T. Arpornthip *et al.* [11] proposed the MOT to be a tool for providing quantitative pressure measurement, however the method is destructive for that it need to turn off ion pump or add more background gas to change the cold atom collision rate with background.

In this paper, we will present a simple, reliable, and no-destructive method for the measurement of vacuum pressure in a magneto-optical trap. It works without specialized equipment but the cold atoms experiment itself. Our experimental results will verify that the presented method is reliable and also applicable for other cold atomic systems. This method is an effective solution to the systems without a measurement or commercial devices.

## 2 Theory

The dynamic of MOT loading process can be described by a rate equation as

$$\frac{dN}{dt} = R - \gamma N(t) - \beta \bar{n} N(t) \quad (1)$$

Where N is the number of atoms in the trap and R is the rate at which atoms are loaded via laser cooling. The trap losses are described by γ, the rate constant for loss due to collisions with all background gases, and β, the rate constant for loss due to inelastic two-body collisions within the trap. The two-body rate also depends on the mean density of the trapped atoms $\bar{n} = \int n(r,t)dr/N$.

The variation of $\bar{n}$ with N is complicated in general. But two typically regimes can be identified depending on the significance of multiple-scattering forces within the MOT [12,13]. For small N, less than of order $10^5$ atoms, the scattering forces are weak and $\bar{n} \approx N(t)/V$ with fixed trap volume V. For larger N, light scattering enforces a constant $\bar{n}$ with $N \propto V$. With the constant atom density approximation, equation (1) can be solved as an exponential loading curve

$$N = \frac{R}{\Gamma}(1 - e^{-\Gamma t}) \quad (2)$$

with the collision loss rate of $\Gamma = \gamma + \beta \bar{n}$. The β mainly depends by the intensity of the trapping laser [14], while γ is insensitive to the intensity of the trapping laser. It can generally be expressed as

$$\gamma = \sum_i n_i \langle \sigma_i v_i \rangle \quad (3)$$

Where the sum is over gas species i in background vacuum, with density $n_i = P_i/k_B T$ according to ideal gas law, gas velocity $v_i$, and collision loss cross section $\sigma_i$. The angle brackets represent an average over the thermal distribution, and the velocity of the trapped atoms can be negligible compared to $v_i$.

The loss cross section $\sigma_i$ is given by [15]

$$\sigma_i = \int_{\theta>\theta_L} \frac{d\sigma}{d\Omega} d\Omega = \int_{\theta>\theta_L} \frac{1}{6}\left(\frac{15\pi}{8}\frac{c_i}{m_i v_i^2}\right)^{1/3} \theta^{-7/3}$$

$$= \left(\frac{15\pi^4}{6}\right)^{1/3}\left(\frac{m_i C_i^2}{m_0 E_i D}\right)^{1/6} \quad (4)$$

Where $d\sigma/d\Omega$ is the differential scattering cross section and $\theta_L = (2m_0 D)^{1/2}/(m_i v_i)$ is the minimum scattering angle required in order to give the target cold atom sufficient energy to escape the trap. D is the MOT trap depths. $m_0$ is the trapped atom mass and $E_i = m_i v_i^2/2$ is the incident energy of the background gases. $C_i$ is the van der Waals coefficient for gas species i and can be estimated using the Slater-Kirkwood formula [16,17].

$$C_i = \frac{3}{2}\frac{\hbar e}{(4\pi\varepsilon_0)^2 m_e^{1/2}}\frac{\alpha_0 \alpha_i}{(\alpha_0/\rho_0)^{1/2} + (\alpha_i/\rho_i)^{1/2}} \quad (5)$$

Where $m_e$ is the electron mass and species i has static electric polarizability $\alpha_i$ and number of valence electrons $\rho_i$. We mark i = 0 to represent the trapped species.

All of the above yields the vacuum pressure to be [18]

$$P_i \approx \frac{(k_B T)^{2/3}}{6.8}\left(\frac{C_i}{m_i}\right)^{-1/3}(Dm_0)^{1/6}\gamma_i \quad (6)$$

Through equation (6), we can get a conclusion that the vacuum pressure is mainly determined by the cold atom collision loss rate constant due to all background gases $\gamma_i$. In fact, the loss due to collisions with all background gases comes mainly from the collisions with the atom species similar to the trapped atoms for homonuclear cold atoms MOT. So equation (6) can be described as

P=kγ  (7)

The proportional coefficient k in equation (7) can be calculated from equation (6). We give the calculated k for common alkali metal and some alkaline earth metal atoms which have been trapped and cooled by MOT in table 1.

| Species | $C_i$ (a.u.) | k (Pa*s) |
|---|---|---|
| Li | 1575.081 | $1.1\times10^{-6}$ |
| Na | 1556.587 | $2.16\times10^{-6}$ |
| K | 3715.314 | $2.11\times10^{-6}$ |
| Rb | 4276.056 | $2.97\times10^{-6}$ |
| Cs | 6052.903 | $3.31\times10^{-6}$ |
| Fr | 4190.761 | $4.85\times10^{-6}$ |
| Yb | 2150.699 | $5.33\times10^{-6}$ |
| Sr | 2998.382 | $3.4\times10^{-6}$ |

**Tab. 1** Calculated proportional coefficient k for common alkali metal and alkaline earth metal, for a 1K trap depth and 300K background gas temperature. The van der Waals coefficient $C_i$ coefficients are in atomic units.

Combing the measurement of cold atom collision loss rate constant γ with the same atoms in background gas and the calculated proportional coefficient k, we can get the MOT vacuum pressure immediately.

## 3 Experimental approach

The cesium atoms are cooled and trapped in a standard vapour-loaded Cs MOT with a background pressure of about $6\times10^{-7}$ Pa. A pair of coils with anti-Helmholtz configuration generates a magnetic gradient of about 12G/cm and three other pairs of coils generate a compensate geomagnetism at the position of cold atom clouds. The trapping and repumping beams are provided by two Littrow external-cavity diode lasers (DL100,Toptica) and locked by saturation absorption spectroscopy technique. With adjusts of acousto-optic modulators (AOMs), the trapping beams are tuned to 15MHz below $6S_{1/2}(F=4)\rightarrow6P_{3/2}(F'=5)$ transition. The repumping beam is tuned to $6S_{1/2}(F=3)\rightarrow6P_{3/2}(F'=4)$ transition. Each trapping beam is 15mm in diameter and 6mW in power. The repumping beam is 15mm in diameter and 5mW in power. In this way, an atomic cloud of $3\times10^{7}$ cesium atoms with a temperature of ~100μk is formed. The density is estimated to $6\times10^{10}cm^{-3}$. The cold atom fluorescence is collected by a convex lens and detected through a Si-Avalanche photodiode module (Licel APD-1.5).

In our experiment, we control the loading process by turning off and on the AOM. Fig. 1 is a typical MOT loading curve while the intensity of each trapping laser beam is 2.3mW/cm$^2$. The fluorescence is measured after the trapping beams are turned on by AOM at time t=0. Data points (black) are the experimentally measured values while the solid (red) curve is a fitting of the data to the exponential form of equation (2). The fitted curve does not agree well with the experimental data in the first second. We believe that this is due to the constant atom density approximation has not been reached under a smaller atomic number. The status is, in fact, the constant atom volume approximation. The fitted curve agrees well with the experimental data from one to six seconds. After six seconds, the fitted curve does not agree again with the experimental data due to formation of steady state. By analyzing the equation (2), we can get a conclusion that the logarithmic plot of the experimental data $-\ln(1-N_t\Gamma/R)$ have a linear relationship with the loading time t under the constant atom density approximation. The inset in Fig. 1 shows the logarithmic plot $-\ln(1-N_t\Gamma/R)$ as a function of the loading time t. The plot is clearly not linear in the first second and after sixth second for the reason mentioned before. We choose the time between the first seconds and the sixth seconds to get the Γ by the linear fitting. These plots are indeed linear, which confirms that trap loading takes place within the regime of constant density as described above. We get the collision loss rate of Γ to be 0.28±0.0002 s$^{-1}$. The uncertainty comes from the fitting.

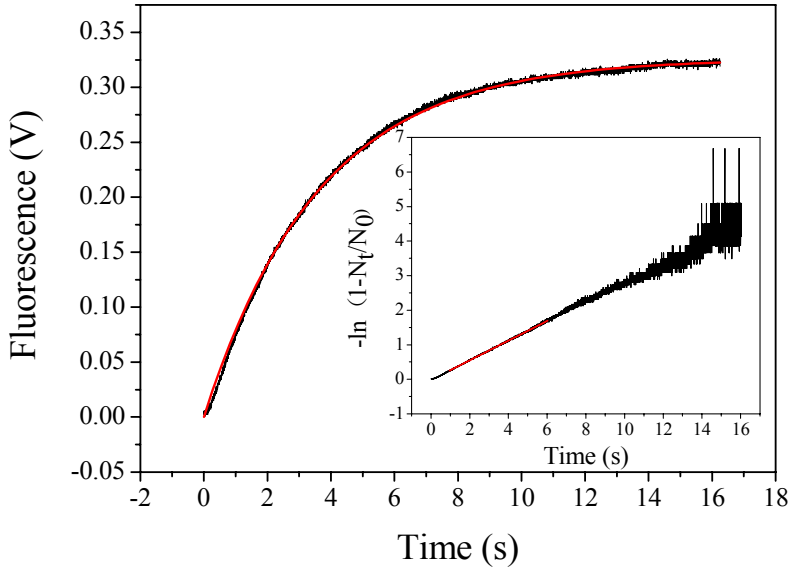

**Fig. 1** A typical MOT loading curve. Inset is the logarithmic plot [−ln(1−NtΓ/R] versus time

## 4 Results and Discussions

As previously mentioned, the rate constant for loss due to inelastic two-body collisions within the trap β mainly are influenced by the intensity of the trapping laser, while the rate constant for loss due to collisions with all background gases γ is insensitive to the intensity of the trapping laser. With the control of the AOM by different modulation voltage, we can generate a series of trapping laser with various intensities. Fig. 2 shows how the collision loss rate Γ vary with the total intensity of trapping lasers. The data points are the experimentally measured values, and the solid line is a linear fitting. The error bars indicate the standard deviations in mean value of up to three times measurements. From the fitting, we can get the value of collision loss rate when the intensity of trapping laser is zero, which is the value of γ. For the Fig. 2, the fitting γ equals to 0.15±0.008 s$^{-1}$. The uncertainty comes from the fitting.

With the proportional coefficient k of 3.31×10$^{-6}$ Pa·s for Cs atoms, we get the vacuum pressure to be (5.0±0.26)×10$^{-7}$ Pa. In order to verify the reliability of this method, we measure the background vacuum pressure by a high precise ionization gauge (937A Gauge Controller). The direct measured value is 5.2×10$^{-7}$ Pa. A good agreement is found for the value derived by our method and the experimental direct measurement. For vacuum pressure measurement, the accuracy of this method is generally acceptable with the comparison to the conventional pressure gauges.

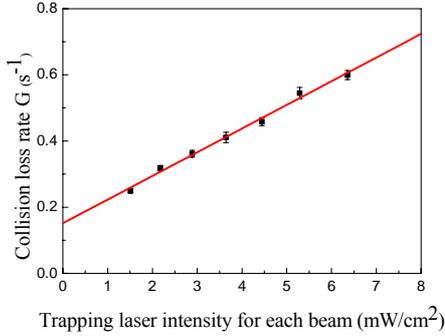

**Fig. 2** The collision loss rate Γ as a function of the trapping laser intensity for each beam.

In order to test the universality of our method, we do a series of experiments by changing the background vacuum pressure. One way is to turn off the ion pump with the result that the vacuum pressure increases quickly, while the vacuum pressure will reduce slowly when the ion pump is re-opened. Another way to vary the background vacuum pressure is to operate the dispenser. Fig. 3 shows the rate constant for loss due to collisions with all background gases γ versus the background vacuum pressure. The data points are the experimentally results while the solid line is a linear fitting. The error bars indicate the standard deviations in mean value of up to three times measurements. From the fitting, we can get the proportional coefficient k to be (3.15±0.14)×10$^{-6}$ Pa*s. The uncertainty comes from the fitting. Our system can do the experimental research about cold Rb atom or Cs atom by the type of source we are using [19]. The reason why fitting value is smaller than the theoretical calculation is considered to be the collision between the Cs atoms and the other Rb atoms background gas. The proportional coefficient k for Cs atoms and Rb atoms is calculated to be the value of 3.02×10$^{-6}$ Pa·s. It can explain the declination of the experimental value from the theoretical value.

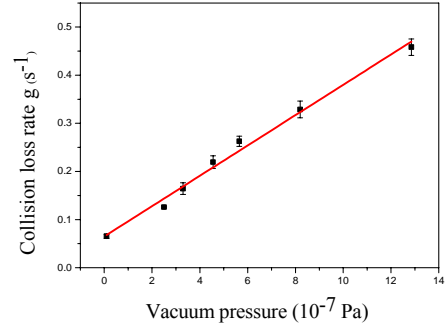

**Fig. 3** The rate constant for loss due to collisions with background gases γ as a function of vacuum pressure.

It is naturally to test the dependence of the proportional coefficient k on varied experimental parameters. It is verified that the proportional coefficient k has no dependence on the trapping laser intensity, detuning, and beam diameters [8]. We now check the dependence of proportional coefficient k on repumping laser intensity with the measured vacuum pressure of 7.2×10$^{-7}$ Pa, which is shown in Fig. 4(a). We also checked the dependence of proportional coefficient k on magnetic field gradient with the measured vacuum pressure of 7.4×10$^{-7}$ Pa, which is shown in Fig. 4(b). The error bars indicate the standard deviations in mean value of up to three times measurements. The proportional coefficient k has no dependence on repumping laser intensity and magnetic field gradient and agrees well with the value of our system 3.15×10$^{-6}$ Pa·s.

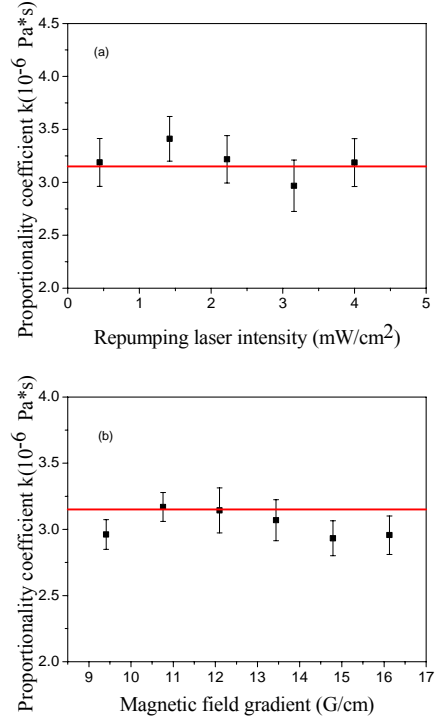

**Fig. 4** The proportional coefficient k as a function of repumping laser intensity (a) and magnetic field gradient (b). The solid line is the proportional

coefficient k of our system with the value of $3.15\times10^{-6}$ Pa·s for Cs atoms.

**5 Conclusion**

In conclusion, we present a simple, reliable, and no-destructive method for the measurement of vacuum pressure in a magneto-optical trap. It works without specialized equipment. The vacuum pressure is calculated by simple ideal gas law. We calculated the proportional coefficient k for common alkali metal and some alkaline earth metal atoms. With the theoretical fitting of the loading curve, we get the collision loss rate Γ. Through the measurement of collision loss rate Γ under different trapping laser intensities, we obtain the rate constant for loss due to collisions with all background gases γ. Then the vacuum pressure P can be obtained by a simple calculation. We also test the dependence of the proportional coefficient k on repumping laser intensity and magnetic field gradient with the result of no influence. The method has the same applies to alkaline earth metal atoms. It meets the requirements of miniaturized and streamlined for business application.

**Acknowledgments** This work was supported by the National Basic Research Program of China (973 Program, Grant No. 2012CB921603), the International Science & Technology Cooperation Program of China (Grant No. 2011DFA12490), the National Natural Science Foundation of China (Grants No. 61275209, 10934004) and NSFC Project for Excellent Research Team (Grant No. 61121064).